\documentclass[
twocolumn, 
secnumarabic,amssymb, preprintnumbers, superscriptaddress, aps, prl]{revtex4-1}
\usepackage{amsmath,times,amssymb,latexsym,graphics,color}
\usepackage{amsmath}

\newcommand{\be}{\begin{equation}}
\newcommand{\ee}{\end{equation}}
\newcommand{\bea}{\begin{eqnarray}}
\newcommand{\eea}{\end{eqnarray}}

\newcommand{\Tr}{\text{Tr}}

\newcommand{\ket}[1]{\left| #1 \right>}

\newcommand{\ba}{\begin{eqnarray}}
\newcommand{\ea}{\end{eqnarray}}

\newcommand{\f}{\frac}

 \def\f {\frac}

\def\Tr{\mathrm{Tr}}
\begin{document}
\preprint{RIKEN-iTHEMS-Report-21}
\title{
Subregion Spectrum Form Factor via Pseudo Entropy}
\date{\today}
\author{Kanato Goto}\email[]{kanato.goto@riken.jp}
\affiliation{\it RIKEN Interdisciplinary Theoretical and Mathematical Sciences (iTHEMS), Wako, Saitama 351-0198, Japan}
\author{Masahiro Nozaki}\email[]{masahiro.nozaki@riken.jp }
\affiliation{\it RIKEN Interdisciplinary Theoretical and Mathematical Sciences (iTHEMS), Wako, Saitama 351-0198, Japan}
\affiliation{\it Kavli Institute for Theoretical Sciences and CAS Center for Excellence in Topological Quantum Computation, University of Chinese Academy of Sciences, Beijing, 100190, China}
\author{Kotaro Tamaoka}\email[]{tamaoka.kotaro@nihon-u.ac.jp}

\affiliation{\it Department of Physics, College of Humanities and Sciences, Nihon University, Sakura-josui, Tokyo 156-8550, Japan}
\begin{abstract}
We introduce a subsystem generalization of the spectral form factor via pseudo entropy, the von-Neumann entropy for the reduced transition matrix. We consider a transition matrix between the thermofield double state and its time-evolved state in two-dimensional conformal field theories, and study the time-dependence of the pseudo entropy for a single interval. We show that the real part of the pseudo entropy behaves similarly to the spectral form factor; it starts from the thermal entropy, initially drops to the minimum, then it starts increasing, and finally approaches the vacuum entanglement entropy. We also study the theory-dependence of its behavior by considering theories on a compact space. 
\end{abstract}
\maketitle
\noindent
\section{Introduction}
Quantum entanglement is being studied extensively in several fields such as high-energy physics, condensed matter physics, and quantum information theory.

In high-energy physics, recent studies in anti-de Sitter (AdS) / conformal field theory (CFT) correspondence \cite{1999IJTP...38.1113M} (AdS/CFT) revealed that information of the geometry in AdS is encoded in the entanglement structure of the dual CFT state, which can be studied using quantum-information-theoretic quantities, such as the entanglement entropy  \cite{Ryu_2006,Ryu_20062}.

In condensed matter physics, quantum-information-theoretic quantities that measure the bipartite and tripartite entanglement have been used to study non-equilibrium phenomena, such as quantum thermalization and information scrambling \cite{Calabrese_2005,Calabrese_2007,Hosur_2016,Nie_2019,Kudler_Flam_2020}.
Moreover, it was found that the spectrum of the reduced density matrix, also called entanglement spectrum, is useful to diagnose quantum phases of matter \cite{Li_2008,Pollmann_2010}. However, the entanglement spectra in quantum field theories are rarely computable analytically \cite{Casini_2011}.

In this paper, we 
introduce a quantity that characterizes 
the entanglement spectra of general subregions. 
In Ref. \cite{2017JHEP...05..118C,Papadodimas_2015}, the following quantities that characterize the energy spectrum have been  proposed 
\begin{align}
 \left|\f{Z(\beta+it)}{Z(\beta)}\right|^2&= \f{1}{Z(\beta)^2}\sum_{n,m}e^{-\beta (E_m+E_n)}e^{i t (E_m-E_n)}\, ,\\
 Z(\beta+it) &\equiv \Tr(e^{-\beta H+it H}),
\end{align}
where $H$ is the Hamiltonian, $\beta$ is the inverse temperature of the system, $E_n$ and $E_m$ are energy eigenvalues, and $Z(\beta)$ is the partition function of the system at $\beta$.
$Z(\beta+it)$ is the  partition function analytically continued to the real time as $\beta\rightarrow \beta +it$. It characterizes the discreteness of the energy spectrum.  It is known that 
the squared quantity $\left|\f{Z(\beta+it)}{Z(\beta)}\right|^2$ 
is  diagnostic of the pair correlation of energy eigenvalues. This squared quantity is called the spectrum form factor (see \cite{mehta2004random}, for example).
The characteristic behavior of the spectrum form factor in a chaotic system is that it initially drops (slope) to the minimum (dip) at the Thouless time, starts increasing (ramp) and finally approaches the constant value (plateau) after the Heisenberg time. The value of spectrum form factor after the Heisenberg time is approximated by the long time-averaged value. 


Here, we propose a subergion generalization of the spectrum form factor by using the pseudo entropy \cite{Nakata_2021}.  We refer the reader to \cite{Chen:2017yzn, Chen:2018hjf, Ma:2020uox} for other generalizations of the spectrum form factor. The pseudo entropy is a generalization of the entanglement entropy to a post-selection process.  Let $|\psi\rangle$ and $|\varphi\rangle$ be a pure state in a bipartite Hilbert space $\mathcal{H}_A\otimes\mathcal{H}_{\bar{A}}$. We introduce a transition matrix,
\begin{equation}
\mathcal{T}^{\psi|\varphi}=\frac{|\psi\rangle\langle\varphi|}{\langle\varphi|\psi\rangle}, \label{transition1}
\end{equation}
and a reduced transition matrix 
\begin{equation}
\mathcal{T}^{\psi|\varphi}_A=\mathrm{Tr}_{\bar{A}}\mathcal{T}^{\psi|\varphi}. \label{transition2}
\end{equation}
Note that (\ref{transition1}) and (\ref{transition2}) are two-state generalizations of the density matrix and the reduced density matrix. In particular, these transition matrices can  appear naturally in the post-selected process, where $|\psi\rangle$ is the initial state and $|\varphi\rangle$ is the final state. Then, we define the pseudo entropy as follows:
\begin{equation}
    S(\mathcal{T}^{\psi|\varphi}_A)=-\Tr_A{\mathcal{T}^{\psi|\varphi}_A\log\mathcal{T}^{\psi|\varphi}_A}. \label{eq:pre}
\end{equation}
In the rest of this paper, we specify the initial and final states as un-quenched and quenched thermofield double (TFD) states

\begin{align} \label{in-out}
\ket{\psi} &= Z^{-\frac{1}{2}}(\beta)\sum_n e^{-\f{\beta}{4} (H_L+H_R)} \ket{n_L}\otimes  \ket{n_R}, \\ ~\ket{\varphi} &=  e^{-\f{it}{2} (H_L+H_R)}\ket{\psi}.\label{in-out2}
\end{align} 
and study the time-dependence of the pseudo entropy in two-dimensional CFTs.
Here the Hamiltonian $H_{L,R}$ acts on the states on $\mathcal{H}_{L,R}$, and $\ket{n_{L,R}}$ is an eigenstate of $H_{L,R}$. In the next section after summary, we will clarify more explicitly the relation between the pseudo entropy for these states and the spectral form factor. The pseudo entropy is complex-valued in general, as the (reduced) transition matrix $\mathcal{T}^{\psi|\varphi}_A$ is in general non-Hermitian. However, the pseudo entropy can be interpreted quantum-information-theoretically as the number of distillable EPR pairs in the post-selection process \cite{Nakata_2021}. Furthermore, the pseudo entropy has a sharp gravity dual in the light of AdS/CFT \cite{Nakata_2021}, and can be used to  detect whether two given states $|\psi\rangle$ and $|\varphi\rangle$
belong to the same quantum phase \cite{Mollabashi:2020yie,Mollabashi:2021xsd,Nishioka:2021cxe}. Throughout this paper, we normalize all dimensionful parameters by a lattice spacing.

\section*{Summary}
We have studied the time evolution of pseudo entropy for a single interval, and our findings can be summarized as follows: 
\begin{itemize}
\item The time evolution of  $\text{Re}[S_A]\equiv\text{Re}[S(\mathcal{T}^{\psi|\varphi}_A)]$ is similar to that of the spectral form factor (see FIG. \ref{fig:pe1}): The time evolution of $\text{Re}[S_A]$ is characterized by two time scales, the {\it subsystem} Thouless time $t_T$ and Heisenberg time $t_H$. In particular, these are characterized by the subsystem size $\ell$ as $t_T\propto\sqrt{\beta\ell}$ and $t_H\propto\ell$ \footnote{We read off the scaling of the Thouless time from the numerical plots of the real part of the pseudo entropy. Small oscillations are observed in the ramp region, and the Heisenberg time is defined as the time at which the pseudo entropy ceases to oscillate.}. Note that the standard Thouless and Heisenberg times in the original spectral form factor are similarly characterized by the total system size $L$, instead of $\ell$.
\item At initial time $t=0$, $\text{Re}[S_A]$ is given by the entanglement entropy of the thermal state. In $0<t<t_T$,  $\text{Re}[S_A]$ decreases. Then, in $t_T<t<t_H$,  $\text{Re}[S_A]$ increases. Finally, in $t>t_H$, $\text{Re}[S_A]$ becomes approximately  constant. The constant value is approximately given by the entanglement entropy of the ground state. In other words, the real part of the pseudo entropy exhibits a dynamical volume-area-law phase transition \cite{Chan:2018upn,Li:2018mcv,Skinner:2018tjl}. 
\item At times   
 considerably larger than the subsystem Heisenberg time, $\text{Re}\left[S_A\right]$ is approximately given by its long-time averaged value. 
\item Our result in the infinite volume system shows the universal behaviour for arbitrary two-dimensional CFTs. We also studied the finite size effects for the two-dimensional critical Ising model and the holographic CFTs. For the critical Ising model, we observed an erratic oscillation after the subsystem Heisenberg time $t_H$, while the holographic CFTs exhibited a self-averaging result.  
\end{itemize}

\section{Spectrum Form Factor from Pseudo Entropy}\label{sec:PEandSSF}
In this section, we elaborate on the connection between the spectrum form factor and pseudo entropy for the TFD states in an explicit example, a two-dimensional CFT. 
By taking the trace of the left Hilbert space $\mathcal{H}_L$, we obtain
\begin{align}
\mathcal{T}^{\psi|\varphi}_R&=\mathrm{Tr}_L\mathcal{T}^{\psi|\varphi}=\dfrac{\mathrm{e}^{-(\beta+it)H_R }}{Z(\beta+it)}. \label{eq:rtm}
\end{align}

The above expression \eqref{eq:rtm} is already suggestive of the spectral form factor if we focus on the real part. We elaborate on this in the following. First, the pseudo entropy for $\mathcal{T}^{\psi|\varphi}_R$ is given by
\begin{align}
S(\mathcal{T}^{\psi|\varphi}_R)&=(\beta+it)\langle H_R \rangle_{\beta+it}+\log Z(\beta+it), \label{eq:per}
\end{align}
where $\langle H_R \rangle_{\beta+it}$ is the thermal expectation value of energy $H_R$ with complexified inverse temperature $\beta+it$.
If we consider a two-dimensional CFT, we obtain an explicit expression for the first term as follows:
\begin{align}
\langle H_R \rangle_{\beta}=\dfrac{c\pi^2}{3}\dfrac{L_R}{\beta^2}, 
\end{align}
where $L_R$ is the volume of the system.
Considering the real part of \eqref{eq:per}, we obtain
\begin{align}
\mathrm{Re} \left[S(\mathcal{T}^{\psi|\varphi}_R)\right]=\dfrac{c\pi^2}{3}\dfrac{\beta L_R}{\beta^2+t^2}+\dfrac{1}{2}\log |Z(\beta+it)|^2,
\end{align}
provides the late-time behavior of the spectrum form factor as the first term is negligible.  

In the following section, we will generalize it to subsystems. Namely, we will further trace over subsystem of $R$ and call the remaining subregion as $A$. In other words, we divide the entire Hilbert space $\mathcal{H}\equiv\mathcal{H}_L\otimes\mathcal{H}_R$ into $\mathcal{H}=\mathcal{H}_A\otimes\mathcal{H}_{\bar{A}}$ such that $A\subset R$. Our main claim is that the pseudo entropy of $\mathcal{T}^{\psi|\varphi}_A$ can be interpreted as a subregion generalization of the spectral form factor.
\section{Pseudo Entropy in a two-dimensional CFT}
\begin{figure}[t]
 \begin{center}
  \resizebox{75mm}{!}{\includegraphics{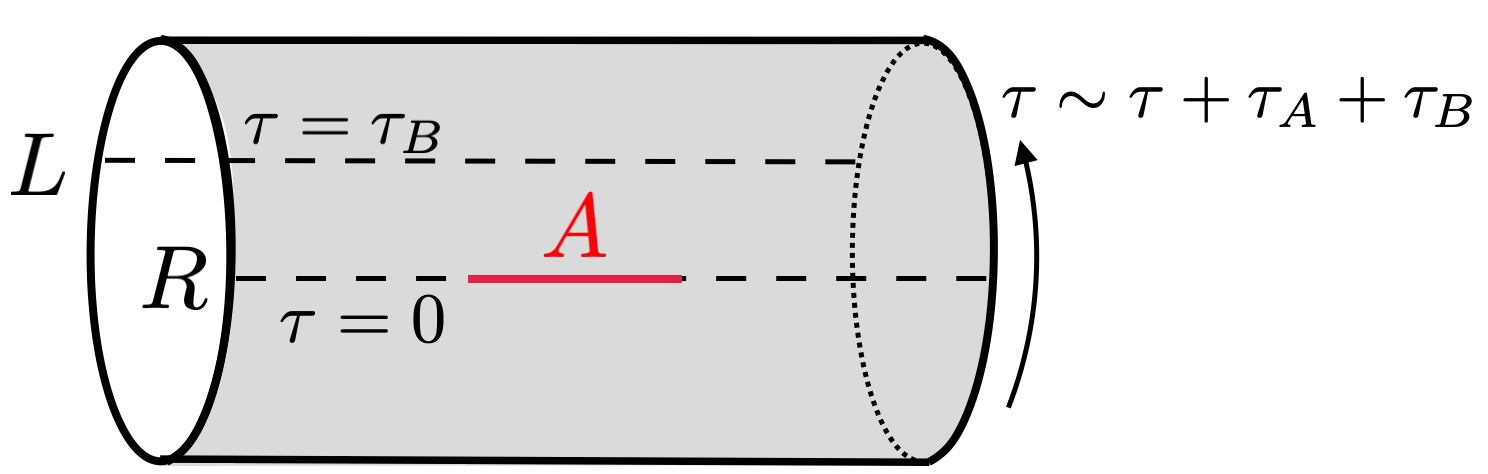}}
 \end{center}
 \caption{The Euclidean path-integral for the reduced transition matrix $\mathcal{T}^{\psi|\varphi}_A$ is performed on the Euclidean cylinder with circumference $\tau_A+\tau_B$ with a cut on A.}
 \label{fig:EPI}
\end{figure}

In this section, we will present a few results regarding the pseudo entropy for the reduced transition matrix
\begin{equation}
\mathcal{T}^{\psi|\varphi}_A=\mathrm{Tr}_{\bar{A}}\mathcal{T}^{\psi|\varphi}.
\end{equation}
Here, we divide the whole system into $A=[X_1,X_2]$ in the right system and its complement.
Using the replica trick \cite{Calabrese:2004eu}, the pseudo entropy can be computed as $n\rightarrow 1$ limit of the (pseudo) R\'enyi entropy, 
\be
S(\mathcal{T}^{\psi|\varphi}_A)=\lim_{n\rightarrow1}S^{(n)}(\mathcal{T}^{\psi|\varphi}_A),
\ee
where we introduced the (pseudo) R\'enyi entropy,
\be
S^{(n)}(\mathcal{T}^{\psi|\varphi}_A)=\dfrac{1}{1-n}\log\mathrm{Tr}_A(\mathcal{T}^{\psi|\varphi}_A)^n.
\ee
To compute the R\'enyi entropy, we first consider the transition matrix $\mathcal{T}^{\psi|\varphi}$  for the initial and the final states $|\psi\rangle,|\varphi\rangle$, prepared by the Euclidean path-integral over $\mathbb{R}\times [0,\tau_A]$ and $\mathbb{R}\times [0,\tau_B]$, respectively
\begin{align}
    |\psi\rangle &=Z^{-\frac{1}{2}}(\tau_A)\sum_n\mathrm{e}^{-\tau_A E_n}|n_Ln_R\rangle,\\
    |\varphi\rangle&=Z^{-\frac{1}{2}}(\tau_B)\sum_n\mathrm{e}^{-\tau_B E_n}|n_Ln_R\rangle.
\end{align}
We will finally perform the Lorentzian continuation $\tau_A=\beta/2$, $\tau_B=\beta/2+it$  after the computations to recover the original initial and the final states in real time.
The reduced transition matrix $\mathcal{T}^{\psi|\varphi}_A$ for  these states are given by the Euclidean path-integral on a cylinder with circumference $\tau_A+\tau_B$ with a cut on $A$ in the right system (see FIG. \ref{fig:EPI}). This is explicitly given by
\begin{align}
\mathcal{T}^{\psi|\varphi}_A&=\mathcal{N}\,{\rm Tr}_{\bar{A}}\sum_{n,m}{e}^{-\tau_A E_n}|n_Ln_R\rangle \langle m_Lm_R|{e}^{-\tau_B E_m}\,,
\end{align}
where $\mathcal{N}=Z(\tau_A+\tau_B)^{-1}$. The R\'enyi entropy can be obtained from $\mathrm{Tr}_A(\mathcal{T}^{\psi|\varphi}_A)^n$ computed by a path-integral for $n$-copies of the cylinder glued along the interval $A$. In the twist operator formalism, this amounts to computing the correlation function of twist operators with dimension $\Delta=\frac{c}{12}(n-1/n)$ on a single copy of the cylinder in an orbifold theory ${\cal T}^n/\mathbb{Z}_n$
\begin{align}
S^{(n)}(\mathcal{T}^{\psi|\varphi}_A)&=\frac{1}{1-n}\log\langle \sigma_n(X_1)\tilde{\sigma}_n(X_2)\rangle\\ \nonumber
&=\frac{1}{1-n}\log\langle \sigma_n(z_1,\bar{z}_1)\tilde{\sigma}_n(z_2,\bar{z}_2)\rangle\, .
\end{align}
Here, we perform the exponential map $z=e^{\frac{2\pi(X+i\tau)}{\tau_A+\tau_B}}$  from the cylinder with circumstance $\tau_A+\tau_B$ to a plane.
Using the conformal symmetry, 
  we can evaluate the two-point function of the twist operators. Finally performing the Lorentzian continuation $\tau_A=\beta/2$, $\tau_B=\beta/2+it$ to recover the original transition matrix in real time and taking the limit $n\rightarrow 1$, we obtain the following form of the pseudo entropy
\begin{equation}\label{pseudo}
    S(\mathcal{T}^{\psi|\varphi}_A)=\frac{c}{6}\log \frac{(\beta+it)^2}{\pi^2}\sinh^2\frac{\pi}{\beta+i t}(X_1-X_2)\, .
    \end{equation}

\begin{figure}
 \begin{center}
  \resizebox{42mm}{!}{\includegraphics{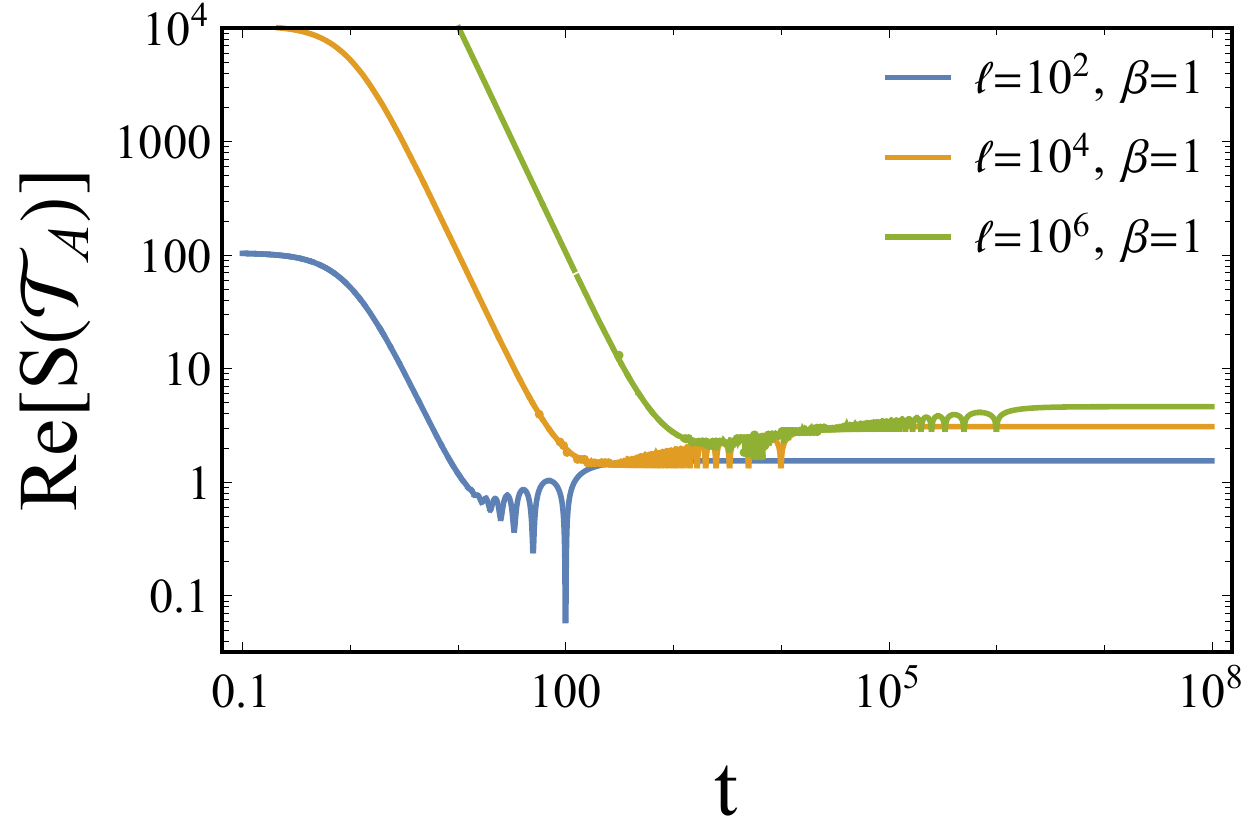}}\;\resizebox{42mm}{!}{\includegraphics{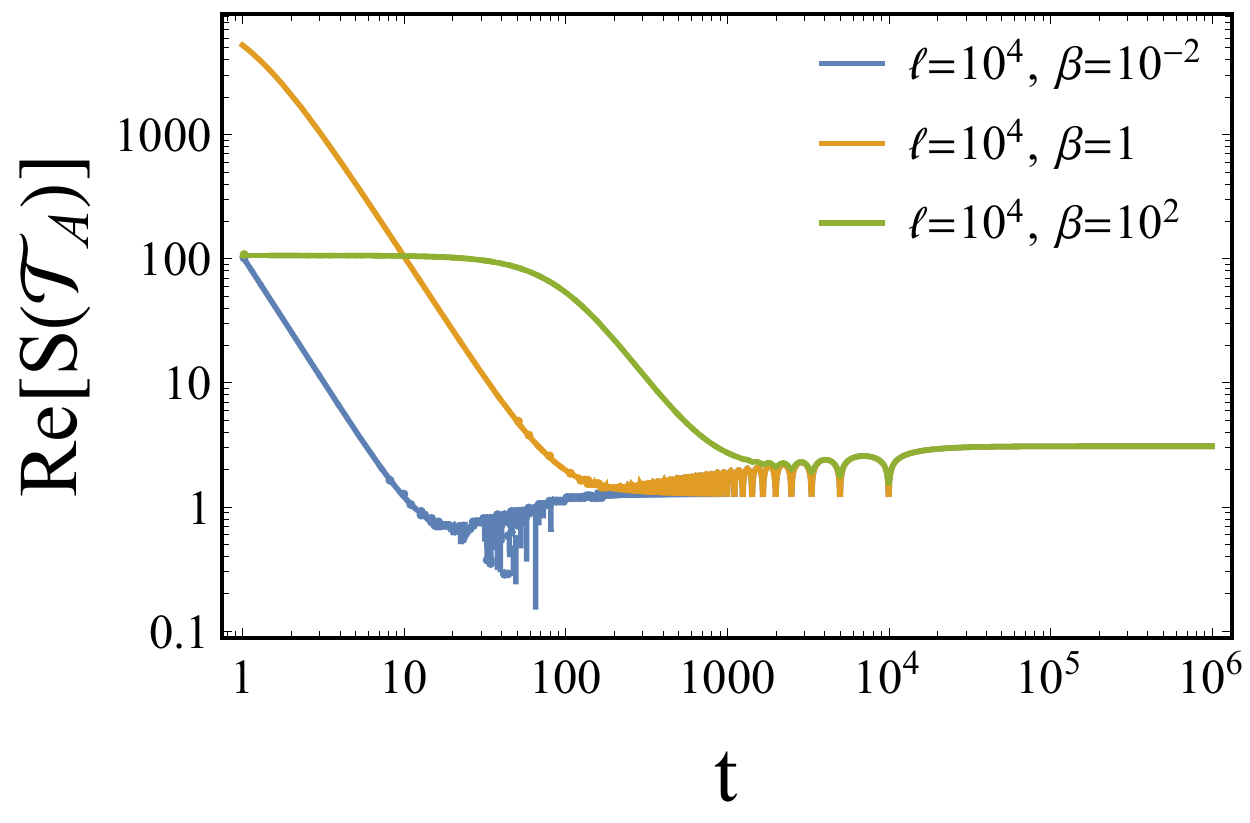}}
 \end{center}
 \caption{Left: The time evolution of the pseudo entropy associated with two states \eqref{in-out} and \eqref{in-out2} exhibits a similar behavior to the spectral form factor. Here, we consider a single interval on $R$ (let us call it $A$) as a subsystem. These log-log plots suggest that $t_T\propto\sqrt{\ell}$ and $t_H\propto\ell$, where $\ell$ is the size of subsystem $A$. Right: The inverse temperature $\beta$ dependence of the pseudo entropy. These plots  suggest that $t_T\propto\sqrt{\beta\ell}$ and $t_H\propto\ell$.}
 \label{fig:pe1}
\end{figure}

    Now, we consider the time dependence of the real part of the pseudo entropy (see FIG. \ref{fig:pe1}).  
At the initial time $t=0$, the pseudo entropy is equal to the entanglement entropy for a single interval in the thermal  state with inverse temperature $\beta$, as expected.  Next, we focus on the late-time  regime $\beta\ll t$, which has relevance to our main interest. Expanding (\ref{pseudo}) in $\beta$, we  obtain
 \begin{equation}
    S(\mathcal{T}^{\psi|\varphi}_A)\sim \frac{c}{6}\log \frac{t^2}{\pi^2}\sin^2\frac{\pi}{t}(X_1-X_2)\, .
    \end{equation}
     at the leading order in $\beta$.
    As we can approximate $\sin\frac{\pi}{t}(X_1-X_2)\sim \frac{\pi}{t}(X_1-X_2) $  in the late-time limit $t\rightarrow \infty$, the pseudo entropy approaches the entanglement entropy for a single interval in the ground state.
     \begin{equation}
    S(\mathcal{T}^{\psi|\varphi}_A)\rightarrow \frac{c}{6}\log (X_1-X_2)^2\, .
    \end{equation}
    Therefore, while the pseudo entropy has complex-values in the intermediate region, it  exhibits a dynamical volume-area law phase transition \cite{Chan:2018upn, Li:2018mcv, Skinner:2018tjl}. From FIG. \ref{fig:pe1}, we can summarize the time-dependence of the real part of the pseudo entropy. The time-evolution of the pseudo entropy starts with the entanglement entropy for the thermal states, and initially drops (slope) to a minimum (dip) at the Thouless time $t_T\propto\sqrt{\beta\ell}$. After that, it starts increasing in time (ramp) with small oscillations. Further, and at the Heisenberg time $t_H=\ell$, it connects to the plateau region. Finally, it approaches the entanglement entropy for the ground state. 
\section{Time average}

In this section, we discuss the late-time behaviors of the pseudo entropy and weak values, {\it i.e.}, the correlation function with respect to transition matrices. In particular, we will see that they are given by the long-time-averaged values of the pseudo entropy and the weak values, and reduce to the entanglement entropy and expectation values for the vacuum state, respectively. 

As a simple example, we start from a two-point function of scalar primary operators in a two-dimensional CFT, 
\begin{equation}
   \mathrm{Tr}\left[\mathcal{T}^{\psi|\varphi}\mathcal{O}(X_1)\mathcal{O}(X_2)\right] = \langle\mathcal{O}(X_1)\mathcal{O}(X_2)\rangle_{\beta+it}.
\end{equation}
Here, the right-hand side is the thermal expectation value of local operators with the complexified temperature $\beta+it$. In other words, we used the same trick as the pseudo entropy. This two point function (on a thermal cylinder) is fixed by the conformal symmetry,
\begin{equation}
\langle\mathcal{O}(X_1)\mathcal{O}(X_2)\rangle_{\beta}=\left|\dfrac{\pi/\beta }{\sinh(\pi (X_1-X_2)/\beta)}\right|^{2\Delta}, \label{eq:2ptwv} 
\end{equation}
where $\Delta$ is the scaling dimension of the scalar operator. Based on these facts, we can easily show
\begin{align}
\lim_{T\rightarrow\infty}\dfrac{1}{T}&\int^T_{0}dt\langle\mathcal{O}(X_1)\mathcal{O}(X_2)\rangle_{\beta+it}\nonumber\\
&=\lim_{t\rightarrow\infty}\langle\mathcal{O}(X_1)\mathcal{O}(X_2)\rangle_{\beta+it}\label{eq:lim_av}\\
&=\dfrac{1}{|X_1-X_2|^{2\Delta}}\label{eq:lim_vac}.
\end{align}
Here, we assumed that $X_1$ and $X_2$ are on the same time slice. The first equality \eqref{eq:lim_av} is obvious from the second one \eqref{eq:lim_vac} as it converges to a time-independent value. Notice that the final expression is given by the vacuum expectation value. This example also explains the late time limit of our pseudo entropy, because it was also given by the two-point function of the twist operators, which are the (scalar) primaries on the orbifold CFT. 


\section{Finite size effects}

\begin{figure}[t]
 \begin{center}
  \resizebox{60mm}{!}{\includegraphics{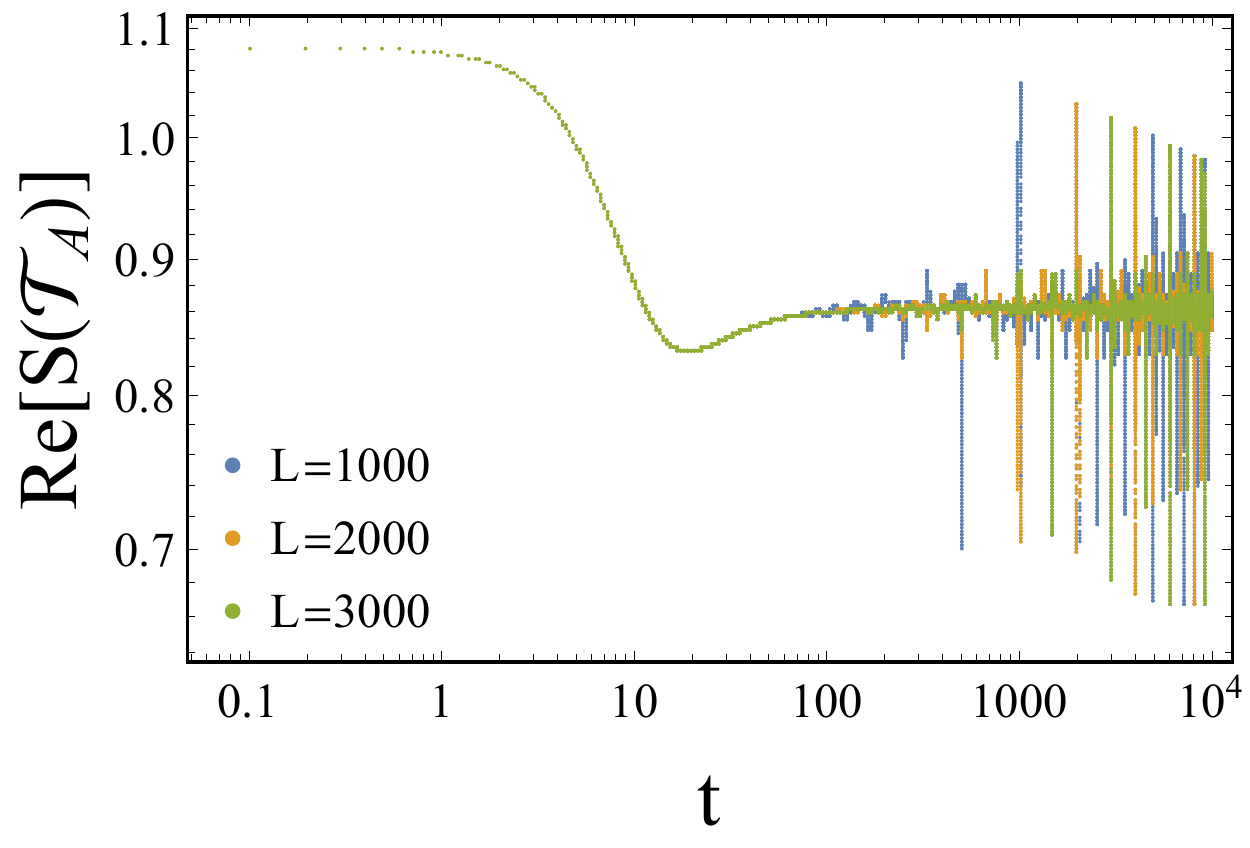}}
 \end{center}
 \caption{Numerical results for critical Ising model with various total system sizes. Here, we set $\beta=10$ and $\ell=10$ for each plot.}
 \label{fig:Ising1}
 \begin{center}
  \resizebox{60mm}{!}{\includegraphics{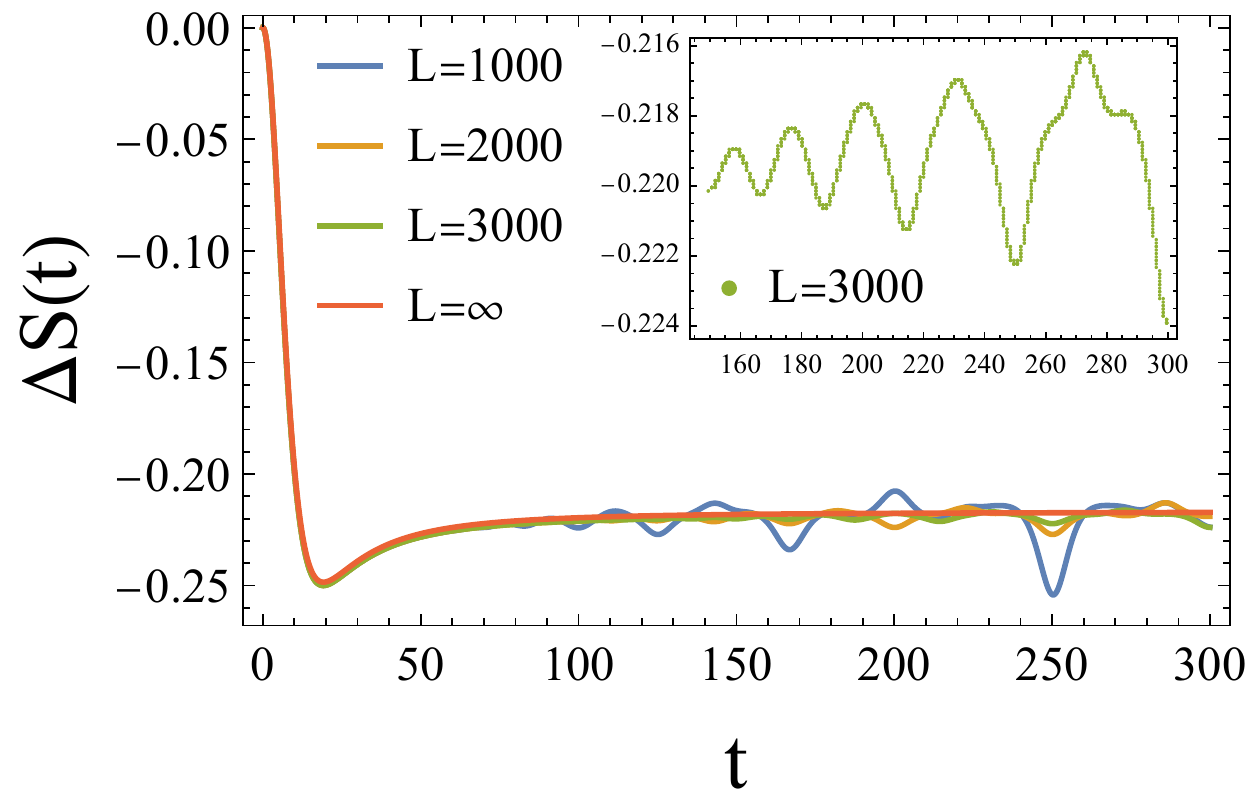}}
 \end{center}
 \caption{Critical Ising model for a finite size system versus that for an infinite size system. Here, we plotted the subtracted pseudo entropy
$\Delta S(t)\equiv \mathrm{Re}[S(\mathcal{T}^{\psi|\varphi}_A)(t)-S(\mathcal{T}^{\psi|\varphi}_A)(0)]$,
for comparison. We set $\beta=10$ and $\ell=10$ for each plot. The red curve corresponds to the analytical result for an infinite-size system. Evidently, as the total system increases, the oscillation becomes smaller. The inset shows smaller erratic oscillations for $L=3000$.}
 \label{fig:Ising2}
\end{figure}

We found that the pseudo entropy for the TFD states behaves as the subsystem generalization of the spectrum form factor. This result is universal for a two-dimensional CFT. Furthermore, the late-time behaviour exactly matches its time-averaging. 
A natural question is why we obtained such a theory-independent result. 
The answer is simply because we considered the entanglement spectrum associated with a single interval in an infinite volume system. In this case, the entanglement entropy provides a universal result for any two-dimensional CFTs (up to the value of the central charge). In other words, the entanglement spectrum is universal \cite{Calabrese_2008}. 

In this section, we consider two finite volume systems where the entanglement spectra of our interest depend on the details of the CFT. 

\subsection{Critical Ising model}
As a concrete example, we study the critical Ising model.
We can numerically study the same pseudo entropy in the finite volume system. 
This can be implemented by the correlator method developed in \cite{Mollabashi:2020yie,Mollabashi:2021xsd}. We plot FIG. \ref{fig:Ising1} and \ref{fig:Ising2}.

In FIG. \ref{fig:Ising1}, we observe that the oscillation becomes more suppressed as the size of total system $L$ increases.  At a very late time after $t\sim L$, we have oscillations with amplitudes close to the initial value. This effect is expected to be a consequence of the fact that the critical Ising model is integrable. We can also confirm such a large amplitude from the analytic expression of R\'enyi entropies \cite{Azeyanagi:2007bj,Herzog:2013py}. 
As a consistency check, we confirmed that for larger systems, we can reproduce the results closer to analytic results in the infinite space, as shown in FIG.  \ref{fig:Ising2}.  

\subsection{Holographic CFT}
On the other hand, based on the holographic formula for pseudo entropy \cite{Nakata_2021}, we can apply the same analytical results as \eqref{pseudo}, when the subsystem is less than half of the total system size. This can be interpreted from the fact that the minimal surfaces do not probe the global structure when the subsystem size is not large \cite{Headrick:2007km, Asplund:2014coa}. 

In this regard, the late-time behaviour of the holographic calculation is indeed self-averaging. This nature of semi-classical gravity is consistent with the original spectral form factor, which has recently attracted significant attention \cite{2017JHEP...05..118C,Saad:2018bqo,Saad:2019lba,Penington:2019kki}.

\section{Discussion}
We proposed a generalization of the spectral form factor to subregions.  Several attempts have been made in this direction previously \cite{Chen:2017yzn, Chen:2018hjf, Ma:2020uox}. In particular, the subergion Thouless and Heisenberg times defined in \cite{Chen:2017yzn, Chen:2018hjf} depend on both the subregion and the total system size. In the future, the origin of this difference should be investigated.  

Here, we only discussed the real part of the pseudo entropy, while the interpretation of the imaginary part remains unclear. An interesting future prospect would be to elucidate the role of the imaginary part.  


We also discussed the holographic formula of pseudo entropy, which yields a self-averaging result, as expected from recent developments in gravity as ensemble averaging \cite{2017JHEP...05..118C,Saad:2018bqo,Saad:2019lba,Penington:2019kki}. Interestingly, this behaviour can be explained by using a single but complexified geometry in the bulk. We would like to stress that this behaviour is also consistent with the standard CFT calculation. Notice that we needed to take into account the non-perturbative effects to probe the ramp and plateau regions in the original spectrum form factor from the holographic analysis. Perhaps such a complex geometry can play a role of an instanton-like solution, which accounts for a certain transition between different spacetime backgrounds, as in our present example. Another interesting future prospect is to investigate the role of such complexified geometries in more general contexts. In particular, it could help to generalize the standard holographic arguments to non-Hermitian systems. In this regard,  it is worth pointing out that the entanglement entropy in non-Hermitian systems can  also be interpreted as a version of the pseudo entropy \cite{Chang:2019jcj}. Therefore, we expect better understanding of pseudo entropy itself would also pave the way in this direction.

\section*{Acknowledgments}
We thank Ali Mollabashi, Shinsei Ryu, Noburo Shiba, Kazuaki Takasan, Tadashi Takayanagi, and Zixia Wei for fruitful discussion. KG is supported by JSPS Grant-in-Aid for Early-Career Scientists 21K13930. MN is supported by JSPS Grant-in-Aid for Early-Career Scientists 19K14724. KT is supported by JSPS Grant-in-Aid for Early-Career Scientists 21K13920.

\bibliographystyle{ieeetr}
\bibliography{reference}

\end{document}